**Probing low temperature non-equilibrium magnetic state in $Co_{2.75}Fe_{0.25}O_{4+\delta}$ spinel oxide using dc magnetization, ac susceptibility and neutron diffraction experiments**

R.N. Bhowmik[*a], Amit Kumar[b,d], A.K. Sinha[c,d], and S.M. Yusuf[b,d]

[a]*Department of Physics, Pondicherry University, R.V. Nagar, Kalapet, Pondicherry 605014, India*

[b]Solid State Physics Division, Bhaba Atomic Research Centre, Mumbai-400085, India

[c]HXAL, SUS, Raja Ramanna Centre for Advanced Technology, Indore- 452013, India

[d]Homi Bhabha National Institute, Anushaktinagar, Mumbai -400 094 India

[*]Corresponding author: Tel.: +91-9944064547; E-mail: rnbhowmik.phy@pondiuni.edu.in

**Abstract**: The low temperature lattice structure and magnetic properties of $Co_{2.75}Fe_{0.25}O_4$ ferrite have been investigated using experimental results from synchrotron x-ray diffraction (SXRD), dc magnetization, ac susceptibility, neutron diffraction and neutron depolarization techniques. The samples have been prepared by chemical co-precipitation of the Fe and Co nitrates solution in high alkaline medium and subsequent thermal annealing of the precipitates in the temperature range of 200- 900 $^0C$. Rietveld refinement of the SXRD patterns at room temperature indicated two-phased cubic spinel structure for the samples annealed at temperatures 200-600 $^0C$. The samples annealed at temperatures 700 $^0C$ and 900 $^0C$ (CF90) have been best fitted with single phased lattice structure. Refinement of the neutron diffraction patterns in the temperature range of 5-300 K confirmed antiferromagnetic (AFM) $Co_3O_4$ and ferrimagnetic (FIM) $Co_{2.75}Fe_{0.25}O_4$ phases for the sample annealed at 600 $^0C$ and single FIM phase of $Co_{2.75}Fe_{0.25}O_4$ for the CF90 sample. Magnetic measurements have shown a non-equilibrium magnetic structure, consisting of the high temperature FIM phase and low temperature AFM phase. The magnetic phases are sensitive to magnetic fields, where high temperature phase is suppressed at higher magnetic fields by enhancing the low temperature AFM phase, irrespective of annealing temperature of the samples.

# 1. INTRODUCTION

The Co rich spinel oxides ($Fe_{3-x}Co_xO_4$; $1<x<3$) have emerged as promising magnetic materials for fundamental study of lattice structure, magnetic spin order and magneto-electronic properties [1-9]. In the formula unit ($AB_2O_4$) of cubic spinel structured Co-Fe oxides, the metal ions (Co, Fe) occupy A (tetrahedral) and B (octahedral) lattice sites with oxygen ions at fcc positions. The heat treatment of nano-structured material and non-equilibrium site distribution of the Co ions in cubic spinel structure play major roles in determining the unconventional lattice structure and magnetic spin order in Co rich spinel oxides [3-9]. In case of $Co_{1.25}Fe_{1.75}O_4$ spinel oxide [10], the high field dc magnetization unusually quenched below 150 K and the neutron diffraction experiment showed an inverse relation of the anomalous lattice expansion with unusual quenching of magnetization at lower temperatures, typically below 150 K, where competition between anisotropy constants and magnetic exchange interactions play important roles. A defect free spinel oxide is expected to be either ferrimagnet (FIM) or antiferromagnet (AFM) if all the divalent cations ($Fe^{2+}$, $Co^{2+}$) occupy the A sites and all trivalent cations (e.g., $Fe^{3+}$, $Co^{3+}$) occupy the B sites [11]. The long ranged magnetic spin order is perturbed under the site exchange of Co and Fe ions and lattice disorder in A and B sites [2, 12]. The non-equilibrium spinel structure, e.g., spinodal decomposition (Co- rich and Fe-rich phases), and its impact on tuning the magnetic and electronic properties have emerged as promising research fields [11-17]. Such materials with inhomogeneous lattice structure and magnetic spin order are interesting for basic understanding of the first order magnetic phase transition, and achieving high magneto-caloric effect and high performance catalytic activities in spinel oxides [12, 18-20].

Despite of showing many unusual magnetic properties in Co rich spinel oxides, there is not much progress on studying their non-equilibrium magnetic properties. We reported the non-equilibrium ferrimagnetic properties of $Co_{2.75}Fe_{0.25}O_4$ spinel oxide at lower temperatures in the range 10-340 K (preliminary information for one sample that was heated at 500 $^0$C [21]) and at higher temperatures (300-950 K) for the samples that were heated at different temperatures [17]. In this work, we report a detailed investigation on the lattice structure and magnetic spin order in the low temperature regime (5-340 K) for extremely high Co rich spinel oxide $Co_{2.75}Fe_{0.25}O_4$ and highlight the role of instability in lattice structure on non-equilibrium magnetic properties.

## 2. EXPERIMENTAL

### A. Sample preparation

The spinel oxide composition $Co_{2.75}Fe_{0.25}O_4$ of approximate amount 2-3 g was prepared through chemical reaction of the required amounts of $Co(NO_3)_2.6H_2O$ and $Fe(NO_3)_3.9H_2O$ nitrate salts. The salts were dissolved in distilled water to yield a transparent aqueous solution at pH ~1.4. The final reaction temperature was maintained at 80 °C for 4h with continuous stirring and pH during chemical reaction was maintained at 11 by adding required amount of NaOH solution. The precipitate was washed several times with distilled water and dried at 100°C. Finally, the resultant powder was heated at 200-250 °C and made free from any significant amount of the bi-product $NaNO_3$. The collected black powder was made into several pellets and annealed for 6 h at temperature 200 $^0$C, 500$^0$C, 700 $^0$C, and 900 $^0$C. The samples were denoted as CF20, CF50, CF70 and CF90, respectively. The heating and cooling rate during annealing of the samples in air was maintained @ 5 °C/min. These samples were used for study of structural properties using synchrotron x-ray diffraction and magnetic properties using dc magnetization ac susceptibility measurements. Similar procedures were followed to prepare the same composition of amount nearly 5 g separately in two batches. One batch (CF60) was annealed at 600 $^0$C for 8 h and second batch (CF90) was annealed at 900 $^0$C for 6 h. These two samples were used for neutron diffraction (ND) and neutron depolarization experiments.

**B. Sample characterization**

The synchrotron X-ray diffraction (SXRD) patterns were recorded at room temperature in the 2θ range of 10-40 ° for the samples CF20, CF50, CF70, CF85 and CF90 using synchrotron radiation facilities at angle dispersive x-ray diffraction beam line 12 of Indus-2, Indore, India. The beam line consists of a Si (111) based double crystal monochromator and two experimental stations, namely, a six circle diffractometer (Huber 5020) with a scintillation point detector and an image plate (Mar 345dtb) area detector. The image plate data were processed using Fit2D program. The wave length was fixed at 0.79990 Å. Photon energy and the sample to detector distance for SXRD were calibrated by using SXRD pattern of LaB$_6$ NIST standard. The dc magnetization was measured using physical properties measurement system (PPMS-EC2, Quantum Design). The real ($\chi^{/}$) and imaginary ($\chi^{//}$) components of magnetic ac susceptibility ($\chi$) were measured for CF50 sample in the frequency range 137-9337 Hz at an ac magnetic field amplitude 10 Oe. Unpolarized ND experiments were carried out in the temperature range 5-300 K using powder diffractometer- I (λ = 1.094 Å) at Dhruva reactor, BARC-Mumbai [22]. A quantity of nearly 4-5 g of powdered material was loaded in a Vanadium-cylindrical can for ND experiments. One-dimensional neutron depolarization measurement was carried out under dc magnetic field of 50 Oe using Polarized Neutron Spectrometer at Dhruva reactor, BARC, Mumbai. Incident neutron beam (λ = 1.205 Å) was polarized in –z (vertically down) direction using single crystal of $Cu_2MnAl$ (111) Heusler alloy. The transmitted neutron beam polarization was analyzed using $Co_{0.92}Fe_{0.08}$ (200) single crystal. The SXRD patterns and ND patterns were modeled by the Rietveld refinement program using FULLPROF Suite software.

## 3. RESULTS AND DISCUSSION

**A. Synchrotron X-ray diffraction**

Fig. 1 shows SXRD pattern of the CF20, CF50, CF70 and CF90 samples, along with fitted data using Rietveld refinement (structural model). The crystalline planes for cubic spinel structure are indexed at the top of Fig. 1(a). The SXRD patterns exclude possibility of any impurity phases, such as $\alpha$-$Fe_2O_3$ and $CoO$ [2, 4, 7]. In the Rietveld refinement, cubic spinel structure of the $Co_{2.75}Fe_{0.25}O_4$ spinel oxide was modeled close to normal spinel structure $[(Co^{2+})_A[Co^{3+}Co^{3+}]_BO_4]$ of $Co_3O_4$ with space group $Fd\bar{3}m$ and Wyckoff positions at A (8a) sites (1/8, 1/8, 1/8) fully occupied by $Co^{2+}$ ions (or minor amount of $Fe^{3+}$ ions), at B (16d) sites (1/2, 1/2, 1/2) co-occupied by $Co^{3+}$ and $Fe^{3+}$ ions, and at 32e sites occupied by oxygen ($O^{2-}$) ions [17, 23]. The occupancies of Co ions at A sites and Fe at B sites were suitably fixed and occupancy of the O ions at 32e sites was allowed either to vary or fixed at 4. The lattice structure was refined using single phase and two-phase models. In the single phase model, the structure was assigned close to $(Co^{2+})_A[Fe_{0.25}^{3+}Co_{1.75}^{3+}]_BO_{4+\delta}$ with occupancy of O ions allowed to vary about 4. In the two phase model, the first phase was modeled to $(Co^{2+})_A[Co^{3+}Co^{3+}]_BO_4$ and the second phase was modeled to $(Co^{2+})_A[Fe_{0.25}^{3+}Co_{1.75}^{3+}]_BO_{4+\delta}$. The structural parameters (lattice constant ($a$), unit cell volume ($V$), oxygen parameter ($u$), Oxygen content, phase fraction and refinement quality factor $\chi^2$) from the Rietveld refinement of SXRD patterns at 300 K are shown in the insets of Fig. 1. We noted that the two-phase model was best fitted for CF20 and CF50 samples, where as the single phase model is best fitted for CF70 and CF90 samples. The lattice constant (*a*) of the samples for single phase model is close to the value obtained for the $Co_3O_4$ phase in two-phase model. The $Co_3O_4$ phase (75-93 %) strongly dominates over the $Co_{2.75}Fe_{0.25}O_{4+\delta}$ phase (25-7 %) in the refinement using two-phase model. The oxygen parameter (*u*) defines a displacement of O atoms with reference to regular tetrahedron at 8a sites and octahedron at 16d sites. The range (0.260-0.263) of *u* values in the samples is consistent to the reported values [17]. A small amount of excess *Oxygen* content (δ) suggests a non-equilibrium lattice structure for the samples with low temperature annealing (200-700 $^0$C) [5, 24], whereas *O* content in the CF90 sample close to normal value 4 implies an equilibrium lattice structure. We calculated grain size of the samples by applying the Debye-Scherrer formula using position and full width at half maximum (FWHM) intensity of (220) and (400) peaks. The grain size of the samples is found in the range 10-25 nm (~10 nm for CF20, 12 nm for CF50, 16 nm for CF70 and 25 nm for CF90).

**B. DC magnetization**

The dc magnetization (M) was measured using conventional zero field cooled (ZFC) and field cooled (FC) modes. In the ZFC mode, the sample was cooled from 300 K to the lowest measurement temperature 10 K in the absence of external magnetic field and a constant magnetic field is applied during recording of magnetization at different temperatures in the range 10-330 K. In the FC measurement, the sample was cooled down from 300 K to 10 K in the presence of magnetic field and magnetization is recorded during increase of the temperature from 10 K to 330 K without changing the cooling magnetic field. The dc magnetization was normalized by the field (H) to get dc magnetic susceptibility (M/H). Fig. 2 (a) shows the temperature dependence of M/H curves in ZFC and FC modes at magnetic field 100 Oe for CF20 sample. The MZFC(T) curve showed two magnetization peaks; a minor peak at temperature $T_{m1}$ ~ 45 K (manifested in the inset of (a)) and an usually strong peak at temperature $T_{m2}$ ~ 270 K, which can be assigned to average blocking temperature of the ferrimagnetic clusters/small particles in Co rich spinel oxide [10]. However, blocking temperature around room temperature and a wide separation between the MFC(T) and MZFC(T) curves below the 330 K is generally unexpected for the spinel oxide whose composition is close to $Co_3O_4$ with AFM transition temperature in the range 30-40 K [3, 23]. The MFC(T) curve shows two peaks; a minor peak at temperature $T_{m1}$ = 20 K and a major peak at temperature $T_{m2}$ = 130 K for applied field 100 Oe (top X-right Y axes in the inset of Fig. 2(a)). The CF50 sample also shows two peaks in the MZFC(T) and MFC(T) curves at 100 Oe (Fig. 2(b)). However, nature of the MZFC(T) and MFC(T) curves differs for the CF50 sample with reference to the CF20 sample. In CF50 sample, the low temperature peak ($T_{m1}$ = 47 K for MZFC and 30 K for MFC) is more prominent than that in high temperature peak ($T_{m2}$ = 205 K for MZFC and 175 K for MFC). One more high temperature peak seems to exist above 330 K for the CF50 sample and it was noted at 375 K from high temperature measurements [17]. The reduction of peak temperatures in the MFC(T) curve in comparison to the MZFC(T) curve at shows magnetic field induced spin order in the samples and it is expected due to anisotropy effect of small particles. The MZFC(T) curves at different magnetic fields have been examined in Fig. 2(c-d). In case of the CF20 sample, magnetization peaks at $T_{m2}$ and $T_{m1}$ both showed a low temperature shift on increasing the field magnitude. The magnitude of dc susceptibility at high temperature regime (>150 K) decreases in contrast to the increase of magnetic susceptibility at low temperature regime (<100 K). In case of the CF50 sample, the peaks at $T_{m2}$ and $T_{m1}$ showed low temperature shift on increasing the magnetic fields, and dc magnetic susceptibility gradually decreased both at high temperature and low temperature regimes, except the susceptibility at the crossing zone (~ 120 K with a dip) has shown an initial increase on increasing the field up to 5 kOe. The magnetization peak at high temperature regime is strongly suppressed in comparison to the low temperature peak. More importantly, the signature of high temperature peak above 300 K is missing at higher magnetic fields ($\geq$ 10 kOe).

The samples (CF70 and CF90), whose SXRD patterns fitted with single phase, showed remarkably different properties in the temperature dependence of dc magnetization curves (Fig. 3 (a-d) with Y axis-log scale and X axis-linear scale). In case of the CF70 sample, MZFC(T) curve at 100 Oe showed a low temperature peak at 72 K, a magnetization dip at about 155 K, and high temperature peak is not visible up to 330 K. In the MFC(T) curve, the low temperature peak shifts to 30 K and magnetization is less sensitive to temperatures in the range 150-330 K. A wide separation is noted between the FC and ZFC magnetization curves throughout the temperatures (10-330 K). The magnetic separation is decreased in case of the CF90 sample, especially around 150 K. The MZFC(T) curve of the CF90 sample showed a low temperature peak at 35 K and the high temperature peak at 300 K. The linear Y-scale plot in the inset of Fig. 3(b) shows that MZFC(T) curve at 100 Oe becomes negative below 37 K and 13 K for the samples CF70 and CF90, respectively. This may be possible as a result of spin compensation between magnetic moments at the A and B sites of magnetically diluted cubic spinel structure [25]. The negative magnetization is not observed in MFC(T) curves, although it decreases below peak temperature $T_{m1}$. As an effect of increasing magnetic field (Fig. 3(c-d)), the magnitude of ZFC susceptibility decreased in both CF70 and CF90 samples except at the crossing zone (~ 150 K) of high temperature and low temperature regimes, where magnetic susceptibility initially increased up to certain value of applied magnetic field (e.g., 2 kOe for CF70 sample and 500 kOe for CF90 sample). At sufficiently higher magnetic fields, the high temperature peak is suppressed and low temperature peak is subsequently enhanced. Such field induced magnetic transformation is very pronounced in CF90 sample. Magnetic field induced variation of the peak temperatures ($T_{m1}$ and $T_{m2}$) is shown in Fig. 4 (a-d). The $T_{m1}$(H) and $T_{m2}$(H) curves follow a power law: T(H) = $a-bH^n$ with exponent (*n*) values in the range 0.14-0.19 and 0.24-0.34 for $T_{m1}$(H) and $T_{m2}$(H) curves, respectively. The fit parameters are shown in Table 2. The validity of such power law with typical exponent values suggests a distribution of size and anisotropy constants of ferrimagnetic clusters in magnetically diluted polycrystalline spinel oxides [26]. The $T_{m2}$(H) values for CF70 and CF90 samples (Fig. 4(c-d)) were not fitted due to limited number of points. Based on the Rietveld refinement of SXRD patterns, we assume two types of clusters; ferrimagnetic clusters dominate at higher temperatures and AFM clusters dominate at lower temperatures. To realize this assumption, we have used Curie-Weiss (C-W) law M/H = $\frac{C}{T-T_0}$ to fit the temperature dependence of dc magnetic susceptibility (M/H) curves at temperatures above $T_{m1}$ and $T_{m2}$, respectively. We used linear fits in H/M vs. T plots [H/M = $\frac{T}{C} - \frac{T_0}{C}$] in the temperature regimes 70-120 K (T > $T_{m1}$) and 260-320 K (T > $T_{m2}$) to calculate the Curie constant (C= $\frac{N\mu^2}{3k_B}$) and the Curie temperature ($T_0$) from the MZFC(T) and MFC(T) curves at 100 Oe (Fig. 4(e-g)) and MZFC(T) curves at 50 kOe (Fig. 4(h-j)). The effective paramagnetic moment per formula unit (µ) and $T_0$ values are shown in Table 1. The calculated parameters may not be true values as the system is not completely free from magnetic exchange interactions in the temperature range that is well below of the real paramagnetic state (> 300 K) [17]. However, it can provide useful information for the nature of local magnetic spin order in the low temperature and high temperature regimes. The fit of MZFC(T) curves at 100 Oe in the low temperature regime shows positive $T_0$, whose strength decreases in the samples with higher annealing temperatures. This is associated with a decrease of effective paramagnetic moment (µ). The fit of MFC(T) curves at 100 Oe provides negative $T_0$ for CF50 sample and it becomes positive with increasing magnitude for the CF70 and CF90 samples. The μ values from MFC(T) curves are relatively large. It shows similar decreasing trend with the increase of annealing temperature of the samples, as noted from MZFC(T) curves. The fit of C-W law in the high temperature regime of MFC(T) curves at 100 Oe for the CF50 and CF70 samples provides unrealistically large values of μ with negative values of $T_0$. On the other hand, the C-W law is found applicable in the low and high temperature regimes of MZFC(T) curves at 50 kOe. In this case also, the μ value has decreased for the sample with higher annealing temperature and the μ values are found relatively large in the high temperature regime. It may be mentioned that positive and negative values of the $T_0$ indicate the dominance of FM and AFM interactions in the spins system. From the analysis of MZFC(T) at 50 kOe curves, we have observed that dominant interactions at low temperature regime changes from AFM ($T_0$ negative) to FM ($T_0$ positive) on increasing the annealing temperature of the samples and dominant interaction seems to be AFM in nature for all the samples at high temperature regime.

The magnetic field dependence of magnetization (M(H)) of the samples has been studied in the field range ±50 kOe at 10 K and 300 K using ZFC mode. The M(H) curves (Fig. 5 (a-d) show features of a ferrimagnet with hysteresis loop and lack of high field magnetic saturation at all temperatures up to 300 K. The insets of Fig. 5 (a, d) clearly show the presence of hysteresis loop at 300 K. The ferrimagnetic spin order at room temperature is technologically interesting by considering the fact that composition of the present spinel oxide is close to $Co_3O_4$ (AFM). In addition to the effect of small amount of Fe content, the two-phase spinel structure also plays an important role in showing high magnetization for the samples with lower annealing temperature. The ferrimagnetic parameters were calculated from the M(H) loops and shown in Table 3. The intercept of M(H) loop on the M axis at H = 0 Oe gives the remanent magnetization ($M_R$). The intercept of M(H) loop on the H axis for M = 0 gives the coercivity ($H_C$). In order to check the high field cooling effect on the ferrimagnetic parameters, we have measured the M(H) loop of the CF20 and CF90 samples at 10 K after field cooling (@50 kOe) from 300 K. Fig. (e-f) has compared the FC and ZFC M(H) loops. It can be seen that FC loop of the CF20 sample (having two-phase structure) has shown a noticeable shift in comparison to the ZFC loop, whereas the CF90 sample (having single phase structure) does not show noticeable shift. The magnitude of exchange bias shift ($H_{exb}$) in the FC loop has been calculated from the difference ($H_{exb} = H_o^{FC} - H_o^{ZFC}$) of the centre ($H_o^{FC} = (H_1^{FC} + H_2^{FC})/2$) of the FC loop with respect to the centre ($H_o^{ZFC} = (H_1^{ZFC} + H_2^{ZFC})/2$) of the ZFC loop, where $H_1$ and $H_2$ are the coercivity on positive and negative magnetic field axis, respectively. The change of coercivity ($\Delta H_c = H_c^{FC} - H_c^{ZFC}$) in the FC loop is calculated with respect to the ZFC loop, where $H_c^{FC} = (|H_1^{FC}| + |H_2^{FC}|)/2$ and $H_c^{ZFC} = (|H_1^{ZFC}| + |H_2^{ZFC}|)/2$. The exchange bias shift ($M_{exb}$) and change in remanent magnetization ($\Delta M_R$) in the FC loop have also been calculated. The CF20 sample has shown a noticeable shift at 10 K along field ($H_{exb}$ = -161 Oe) and magnetization ($M_{exb}$ = +0.05 emu/g) directions, along with a large in coercivity ($\Delta H_c \sim$ +687 Oe) and remanent magnetization ($\Delta M_R \sim$ +0.1639 emu/g). This is an effect of heterogeneous spin structure, having AFM and ferrimagnetic spin orders, in the nano-structured samples that were prepared by low temperature annealing [24, 27-28]. We have measured M(H) curves at different temperatures for the CF50 sample to study the effect of temperature variation on ferrimagnetic parameters. Some of the M(H) curves are shown in Fig. 6 (a-e). The initial M(H: 0 → 70 kOe) curves at high fields (> 20 kOe), as shown in Fig. 6(f), were fitted using the law of approach to saturation of magnetization: $M(H) = M_{sat}\left(1 - \frac{A}{H} - \frac{B}{H^2}\right) + \chi_p H$ to derive saturation magnetization ($M_{sat}$), effective magneto-crystalline anisotropy constant ($K_{eff}$) using the relation $B = (8/105)(K_{eff}/M_{Sat})^2$, and paramagnetic susceptibility ($\chi_p$). Table 3 compared the values of magnetic parameters at 10 K and 300 K for CF20, CF50, CF70 and CF90 samples. At 10 K, the $H_c$ and $\chi_p$ of the material have gradually decreased, where as an increasing trend with intermediate fluctuation of the parameters $M_R$, $M_{sat}$ and $K_{eff}$ has been noted with the increase of annealing temperature of the material. These ferriparameters have shown finite value at 300 K. In the temperature variation of the ferrimagnetic parameters of CF50 sample (Fig. 6 (g-i)), the $M_R$ has gradually decreased on increasing the temperature from 10 K to 300 K; whereas other parameters ($M_{sat}$, $H_c$, $K_{eff}$ and $\chi_p$) have shown a non-monotonic temperature variation with a local maximum in the range 50-175 K. The results differentiate the anisotropy constant at high temperatures (>100 K) from low temperatures (< 100 K). It affects on the variation of saturation magnetization of the sample and shows the existence of clusters with different spin orders [26].

**C. AC susceptibility**

The existence of clusters with different spin orders (non-equilibrium magnetic state) in the CF50 sample is illustrated by the temperature dependence of real ($\chi'$) and imaginary ($\chi''$) components of ac susceptibility (Fig. 7). The ac susceptibility was measured at ac field amplitude 10 Oe with driving frequency in the range 137-9337 Hz. The $\chi'$ (T) curves showed the existence of two peaks at $T_{m1}$ and $T_{m2}$ within the temperature range 10-300 K and a possible third peak above 300 K. The $\chi''$ (T) curves showed the existence of three peaks corresponds to the peaks in $\chi'$ (T) curves. The non-zero value of $\chi''$ (T) curves corresponds to the peaks in $\chi'$ (T) curves suggests the ferrimagnetic spin order in the system; rather than AFM order where imaginary ($\chi''$) part is expected to be nearly zero. The peak temperatures $T_{m1}$ and $T_{m2}$ observed in both $\chi'$(T) and $\chi''$(T) curves appears to be independent of driven frequencies. It rules out typical spin glass or cluster glass freezing at the peak temperatures $T_{m1}$ and $T_{m2}$ [29-30]. It is interesting to see that low temperature peak in the $\chi'(T)$ curves occurs at lower position ($T_{m1} \sim 43$ K) than that in the $\chi''(T)$ curves ($T_{m1} \sim 50$ K). This is in contrast to the peak positions at $T_{m2}$, where peak in the $\chi''(T)$ curves occurs at lower temperature ($T_{m2} \sim 190$ K) than that in the $\chi'(T)$ curves ($T_{m2} \sim 205$ K). The presence of a frequency independent small kink at 33 K in the $\chi'(T)$ curves suggests a secondary AFM phase of $Co_3O_4$ [23]. The magnitude of the $\chi''(T)$ curves also does not show any significant change or peak corresponding to the kink in $\chi'(T)$ curves, indicating a typical AFM transition temperature. The substantially higher value of the peak temperature (at 43 K) for the primary phase suggests that it is the small amount of $Fe^{3+}$ doped $Co_3O_4$ phase. The frequency independent peak in the $\chi''(T)$ curves corresponding to the peak temperature at 43 K in the $\chi'(T)$ curves indicates that it is the low temperature AFM transition temperature for the Fe doped $Co_3O_4$ phase. On the other hand, a considerable frequency shift in the dip of $\chi'(T)$ curves at $\sim$ 250 K (which is associated to a peak in the $\chi'(T)$ curves above 300 K) and in the shoulder of corresponding $\chi''(T)$ curves suggest superparamagnetic blocking of the ferrimagnetic clusters at higher temperatures. The freezing of the associated high temperature AFM clusters is occurred at $T_{m2} \sim 205$ K. The ac susceptibility data show a local distribution of magnetic spin order; giving rise to clusters of ferrimagnetic and AFM spin order, and splitting of the low temperature peak confirms a distribution of atomic composition in lattice structure [3, 20, 29].

### D. Neutron diffraction and neutron depolarization

We have performed ND and neutron depolarization experiments for two (CF60 and CF90) samples to confirm the magnetic phase, site distribution of metal ions and their magnetic moment. Fig. 8(a-d) shows the Rietveld refinement of the ND patterns at 5 K and 300 K. The observed peaks, as indexed for ND pattern at 5 K for CF90 sample, matched to cubic spinel structure without any impurity phase. Consistent with modeling of the SXRD patterns at 300 K, the ND patterns of CF90 sample is matched with cubic spinel structured (space group $Fd\bar{3}m$) single phase of $Co_{2.75}Fe_{0.25}O_4$, whereas ND patterns of the CF60 sample is accompanied with secondary phase of $Co_3O_4$ (may be doped with minor amount of Fe ions). The identical crystal symmetry (chemical and magnetic unit cells) was used in Rietveld refinement, by considering the overlapping of nuclear and magnetic contributions to the Bragg peaks [10]. The $u$ values from refinement of ND patterns are found in the range 0.262-0.264 and consistent to the values from Rietveld refinement of the SXRD data at 300 K and high temperatures [17]. The Bragg peaks ((111), (220) and (222)) at lower scattering angles largely contribute to magnetic structure and the Bragg peaks ((311), (333), (533), (553), (800)) at higher scattering angles contribute mainly to nuclear (lattice) structure. The peaks (111) and (400) have magnetic contributions from the both A and B sites, whereas the peaks (220) and (222) represent magnetic contributions only from A sites and B sites, respectively. Temperature variation of the intensity of (220), (222) and (400) peaks (not shown in graph) indicated a dominant role of A site moment in determining the net moment in cubic spinel structure. A clear visibility of the (200) peak at about 15.45 $^0$ (Bragg peak only from magnetic phase) in CF60 sample shows the magnetic ordering of $Co_3O_4$ phase, which is absent in CF90 sample. In Fig.9 (a-f), we have shown the temperature variation of the structural and magnetic parameters (lattice parameter ($a$), phase fraction, magnetic moments) obtained from Rietveld refinement of the ND patterns of the CF60 and CF90 samples. The lattice parameter of the CF60 sample has shown a slow increment above 50 K for both the phases ($Co_3O_4$ and $Co_{2.75}Fe_{0.25}O_4$) and a fast decrease is noted below 50 K. The CF90 sample has shown a non-linear lattice expansion with temperature. In the two-phased structure of CF60 sample, the $Co_3O_4$ phase (phase 1) is nearly 63 % and $Co_{2.75}Fe_{0.25}O_4$ phase (phase 2) is 37 % for temperatures ≥ 100 K. The phase 2 is sharply increased below 100 K at the expense of phase 1. As shown in the inset of Fig. 9 (c) for CF60 sample, magnetic contribution of the phase 1 has sharply decreased from a finite value (5 %) at 5 K to nearly zero at 25 K and absolutely zero at temperatures ≥ 50 K, whereas magnetic contribution of the phase 2 slowly decreased from 4% at 5 K to 2 % at 300 K. In the single phase ($Co_{2.75}Fe_{0.25}O_4$) structure of CF90 sample, the magnetic contribution is nearly 3.5 % and nuclear phase is nearly 96.5 % at 5 K. The magnetic fraction decreased towards 0 % by increasing the nuclear fraction in lattice structure close to 100 % at 300 K. The temperature variation of the intensities of (220), (222) and (400) peaks (nor shown) and refinement results of the magnetic structure indicated that net moment per formula unit in both the samples is dominated by the A site contribution and magnetic moment in the CF60 sample is higher than that in the CF90 sample. It appears that net moment in both the samples is largely controlled by the B site moment at temperatures ≤ 50 K, where net moment in the CF60 sample increases unlike a decrease in the CF90 sample. There is a fluctuation in the calculation of magnetic moment at higher temperatures for the CF90 sample. The accuracy of the calculated moments for temperatures above 200 K is poor, most probably due to low moment values. Interestingly, the moment values per formula unit are ~2.89 $\mu_B$ and 1.47 $\mu_B$ for the A and B sites, respective at 25 K for the CF60 sample. These values are close to spin only contribution of 3 μB for $Co^{2+}$ ion (high spin state) at the A site, and for 0.25 numbers of $Fe^{3+}$ ion with moment 5 μB (high spin state) and remaining number (~1.75) of $Co^{3+}$ ions having 0 μB (low spin state) at the B sites [2-3]. On the other hand, moment values per formula unit of the CF90 sample are ~1.16 $\mu_B$ and 1.35 $\mu_B$ for the A and B sites, respectively at 25 K. Although chemical composition of our samples is supposed to be close to the distribution of cations (($Co^{2+}$)$_A$[$Fe^{3+}_{0.25}Co^{3+}_{1.75}$]$_B$$O_4$), but exchange of a small amount of $Fe^{3+}$ and $Co^{2+}$ ions between A and B sites, especially in the samples that were prepared by annealing at lower temperatures, can show higher magnetic moment in comparison to CF90 sample. The defective spinel structure (chemical formula unit is $A_{1-x1}B_{2-x2}[]_yO_4$ or $AB_2O_{4+\delta}$) is another factor that can contribute to higher magnetic moment and properties [16-17, 19, 23]. In fact, the refinement of SXRD patterns for the samples with lower annealing temperatures indicated excess oxygen ($\delta \sim 0.374$ for CF20 sample) in the formula unit of spinel structure. Since chemical composition of the CF90 sample is close to the equilibrium structure of $Co_{2.75}Fe_{0.25}O_4$, its low temperature magnetic properties are not much affected by the lattice defects. But, chemical clustering/heterogeneity in the lattice structure can introduce an atomic level distribution in cation's order, anisotropy, magnetic exchange interactions and magnetic moment in the A and B sites of Co rich spinel oxides [16, 18, 31-32].

The neutron depolarization experiment was performed to correlate the variation of domain magnetization of the CF60 and CF90 samples with the properties from low temperature dc magnetization. Fig. 10 shows the temperature dependence of the transmitted part of polarized incident neutron beam after passing though the sample. The depolarization of incident neutron beam in case of the CF60 sample is relatively weak in comparison to the CF90 sample. In the temperature range 110-300 K, a weak signature of depolarization effect is observed for the CF90 and the depolarization effect becomes stronger on lowering the temperature below 110 K. Keeping in mind the temperature dependence of low field (100 Oe) magnetization curve, it is suggested that a strong depolarization below 110 K is associated with the onset of a strong ferrimagnetic spin ordering in the system. Interestingly, the increase of transmitted polarization (decrease of depolarization) below 50 K may be correlated to a reversing of spin order into AFM state or reduction of ferromagnetic domain size [33-35]. The reduction of FM interactions and increase of AFM interactions leads to a competition between the magnetic ordering among the A and B site moments. This could be the reason of the negative magnetization observed at small applied field (100 Oe) at temperatures below 50 K. Hence, the depolarization experiment reveals the coexistence of ferrimagnetic and AFM spin ordering in the present spinel oxide samples.

## 4. CONCLUSIONS

We have studied the low temperature lattice structure and magnetic properties of the Co spinel oxide with chemical composition $Co_{2.75}Fe_{0.25}O_{4+\delta}$ ($\delta \leq 0.374$). Grain size of the samples has been found in the range 10-25 nm. The material has shown non-equilibrium lattice structure for the samples with low annealing temperature ($\leq 600\ ^0C$) and equilibrium lattice structure for the sample with annealing temperatures at 700 $^0C$ and 900 $^0C$. All the samples have shown signatures of multiple magnetic transitions (at least two transitions below 300 K). The transition temperatures are magnetic field sensitive. The field response depends on the magnetic spin order at low temperature regime and high temperature regime. The magnetization corresponding to high temperature phase is suppressed at higher magnetic fields, unlike the dominance of magnetization at low temperature phase. Overall, the magnetic features are controlled by the randomly distributed ferrimagnetic and AFM clusters, where the AFM clusters control the low temperature magnetic phase and the ferrimagnetic clusters control the high temperature magnetic phase. The coexistence and completion between ferrimagnetic and AFM interactions introduce a non-equilibrium magnetic state in the samples. The non-equilibrium magnetic state observed in the low temperature range 10-300 K(present work) is less affected by defective spinel structure and different from the defect induced non-equilibrium magnetic state in the high temperature range 300-950 K (demonstrated in Ref. [17]).

**ACKNOWLEDGMENTS**

RNB acknowledges the Research grant from UGC-DAE- CSR (No M-252/2017/1022), Gov. of India. RNB also thanks CIF, Pondicherry University for magnetic measurements, RRCAT-Indus 2, Indore for synchrotron x-ray diffraction study and Bhabha Atomic Research Centre (BARC), Mumbai for neutron diffraction and depolarization experiments.

Table 1. Magnetic parameters obtained from fit of the Curie-Weiss law to the temperature dependence of magnetization in low temperature (LT) and high temperature (HT) regimes of the MZFC and FC curves.

| Sample | 100 Oe (ZFC) | | 100 Oe (FC) | | | | 50 kOe (ZFC) | | | |
|---|---|---|---|---|---|---|---|---|---|---|
| | LT | | LT | | HT | | LT | | HT | |
| | $\mu_B$ | $T_0$ (K) | $\mu_B$ | $T_0$ (K) | $\mu_B$ | $T_0$ (K) | $\mu_B$ | $T_0$ (K) | $\mu_B$ | $T_0$ (K) |
| FCF50 | 3.0975 | 6064 | 21.341 | - 60 | 25.642 | - 33 | 4.993 | - 60 | 5.978 | - 114 |
| CF70 | 1.289 | 131 | 6.145 | 77 | 27.309 | - 1076 | 3.659 | -3 | 5.609 | - 225 |
| CF90 | 1.565 | 95 | 3.348 | 84 | -- | -- | 2.200 | 66 | 3.844 | - 164 |

Table 2. Parameters from fitting of the peak temperatures ($T_{m1}$ and $T_{m2}$) of MZFC(T) curves at different magnetic fields (H) using the power law $T_m(H) = a - b*H^n$ for four samples.

| Sample | $T_{m1}(H)$ data | | | $T_{m2}(H)$ data | | |
|---|---|---|---|---|---|---|
| | $a$ (K) | $b$ (K/Oe$^n$) | $n$ | $a$ (K) | $b$ (K/Oe$^n$) | $n$ |
| CF20 | 59.38 ± 2.08 | 6.45 ± 0.47 | 0.19 ± 0.01 | 327.05 ± 18.63 | 18.38 ± 8.68 | 0.24 ± 0.06 |
| CF50 | 59.59 ± 1.30 | 7.24 ± 0.68 | 0.16 ± 0.01 | 231.41 ± 11.08 | 2.78 ± 1.02 | 0.34 ± 0.16 |
| CF70 | 97.25 ± 2.73 | 15.28 ± 3.49 | 0.17 ± 0.02 | -- | -- | -- |
| CF90 | 45.35 ± 1.60 | 6.17 ± 1.50 | 0.14 ± 0.32 | -- | -- | -- |

Table 3. Magnetic parameters from M(H) loop and initial M(H) curves at 10 K and 300 K for different samples.

| Sample | 10 K (ZFC) | | | | | 300 K | | | | |
|---|---|---|---|---|---|---|---|---|---|---|
| | $H_C$ (Oe) | $M_R$ (emu/g) | $M_{sat}$ (emu/g) | $K_{eff}$ (Oe - emu/g) | $\chi_p$ ($10^{-5}$ emu/g/Oe) | $H_C$ (Oe) | $M_R$ (emu/g) | $M_{sat}$ (emu/g) | $K_{eff}$ (Oe - emu/g) | $\chi_p$ ($10^{-5}$ emu/g/Oe) |
| CF20 | 7930 | 1.8134 | 3.9529 | 11108 | 6.5379 | 40 | 0.1312 | 2.2893 | 34895 | 2.5159 |
| CF50 | 4920 | 1.2842 | 3.9306 | 20384 | 6.914 | 192 | 0.0890 | 0.9369 | 7055 | 2.74 |
| CF70 | 2980 | 1.5330 | 5.0005 | 95029 | 4.5894 | 500 | 0.1390 | 0.6212 | 54204 | 1.9153 |
| CF90 | 790 | 1.27 | 7.8655 | 91380 | 4.398 | 40 | 0.0014 | -- | -- | -- |

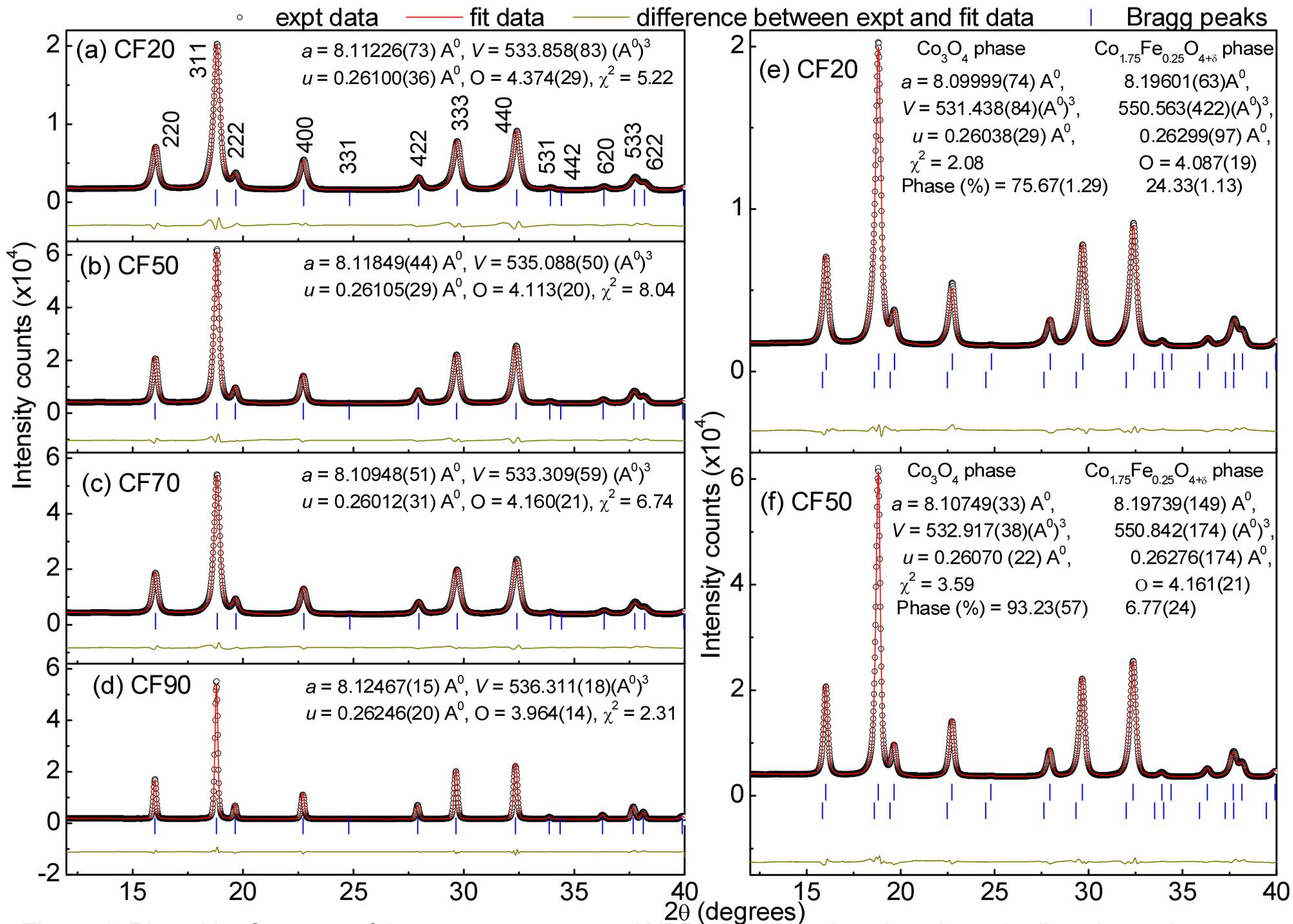

Figure 1. Rietveld refinement of the room temperature SXRD patterns using sing phase (a-d) and two phase models (e-f) with fit parameters shown in the insets. Bragg lines for cubic structure are indicated in (a).

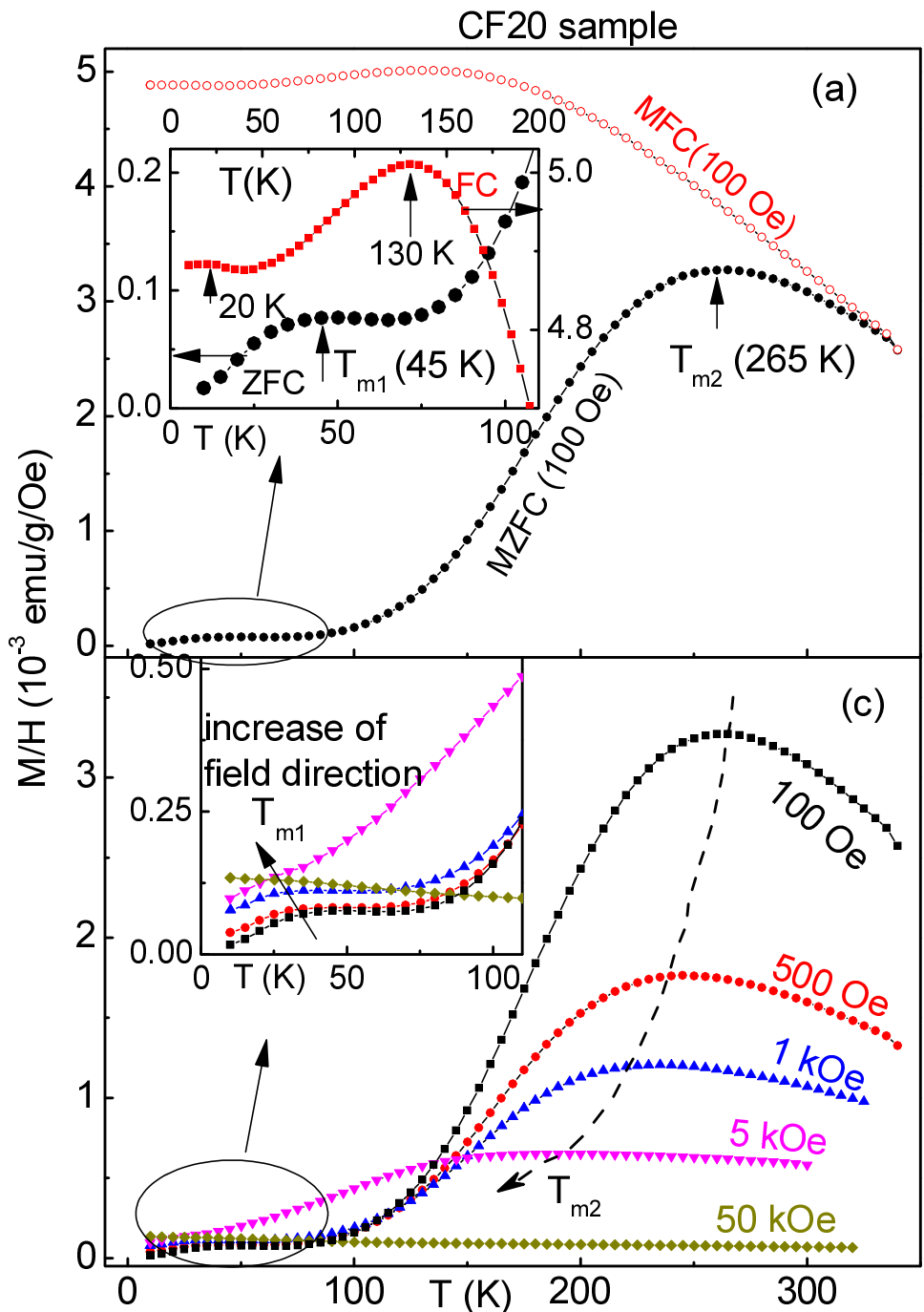
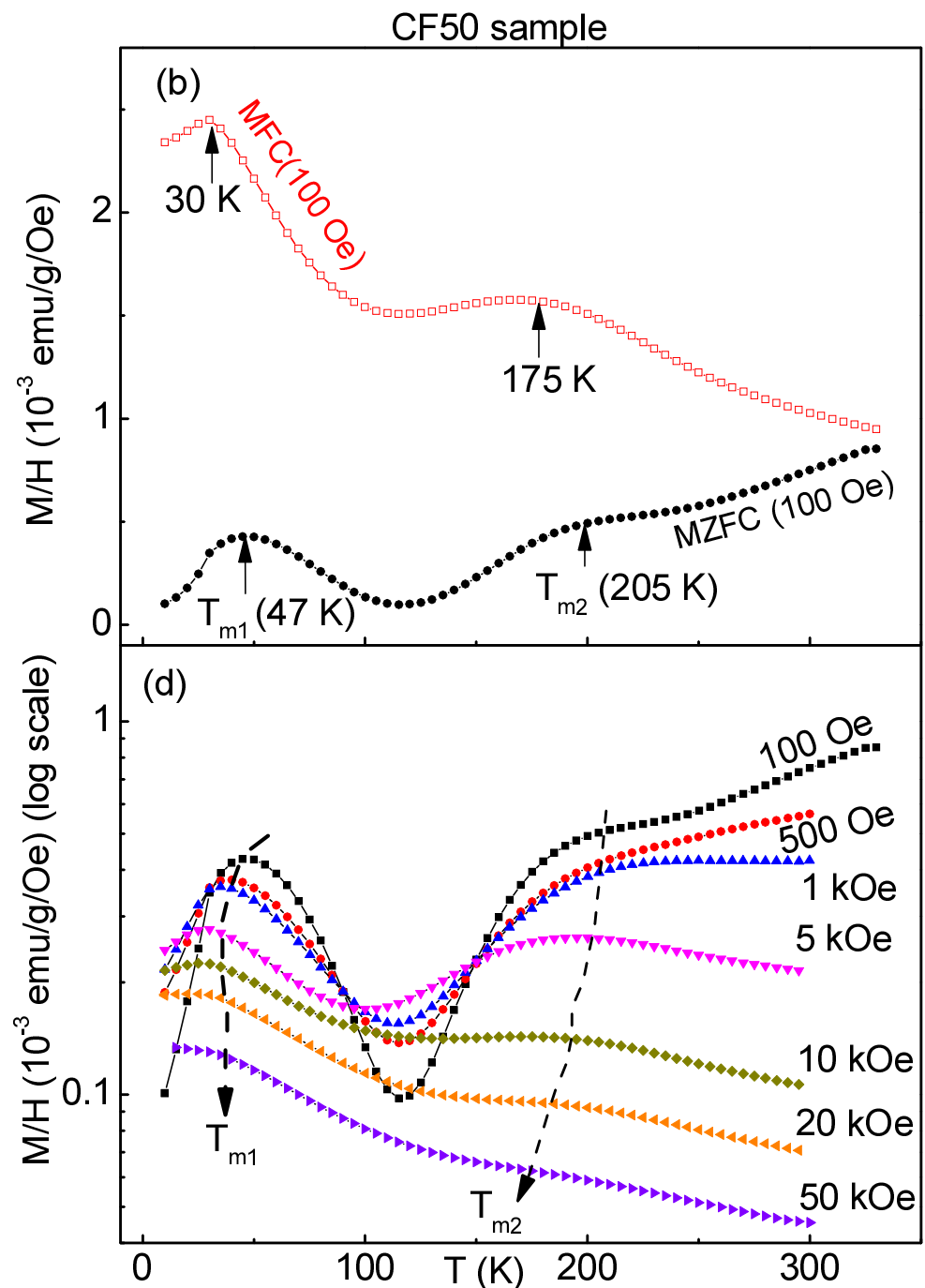

Figure 2. The MZFC(T) and MFC(T) curves at 100 Oe (a-b) and MZFC(T) curves at higher magnetic fields (c-d) for the CF20 and CF50 samples. The magnetic blocking/freezing temperatures ($T_{m1}$ and $T_{m2}$) are marked by vertical arrows. The magnetic field dependence of low temperature peak of CF20 sample is shown in insets of a and c.

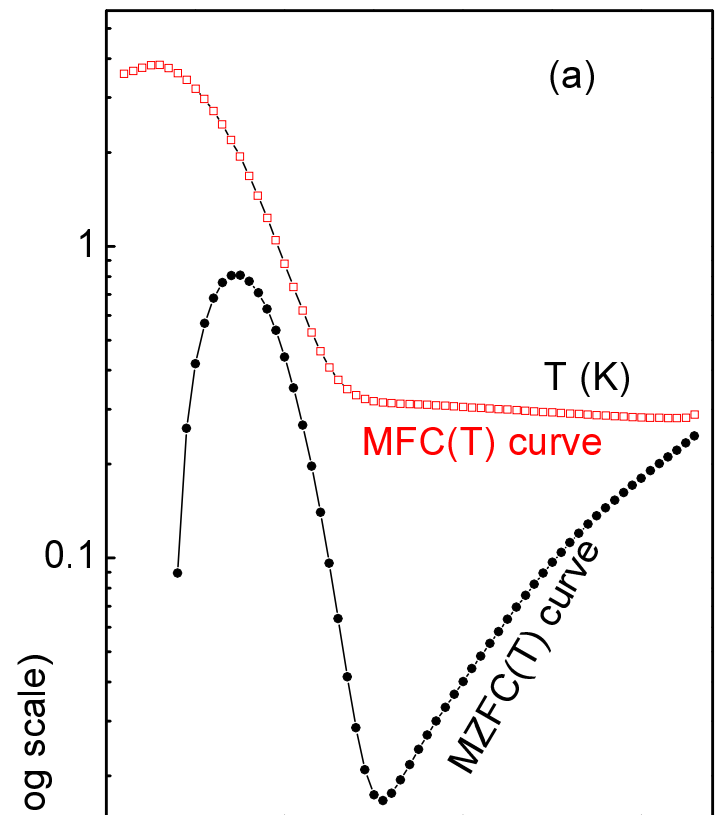
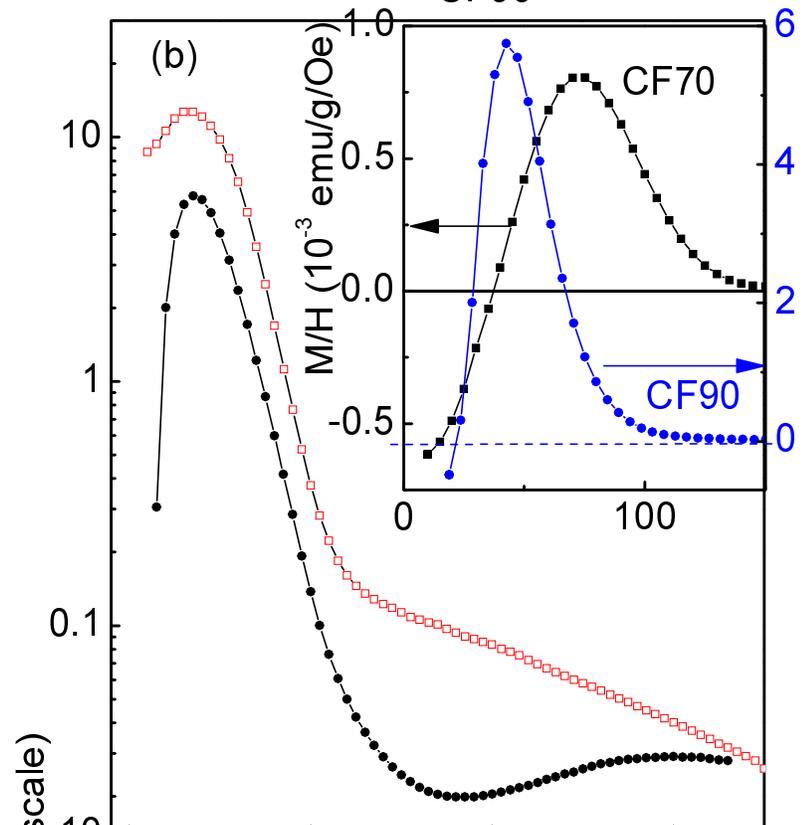
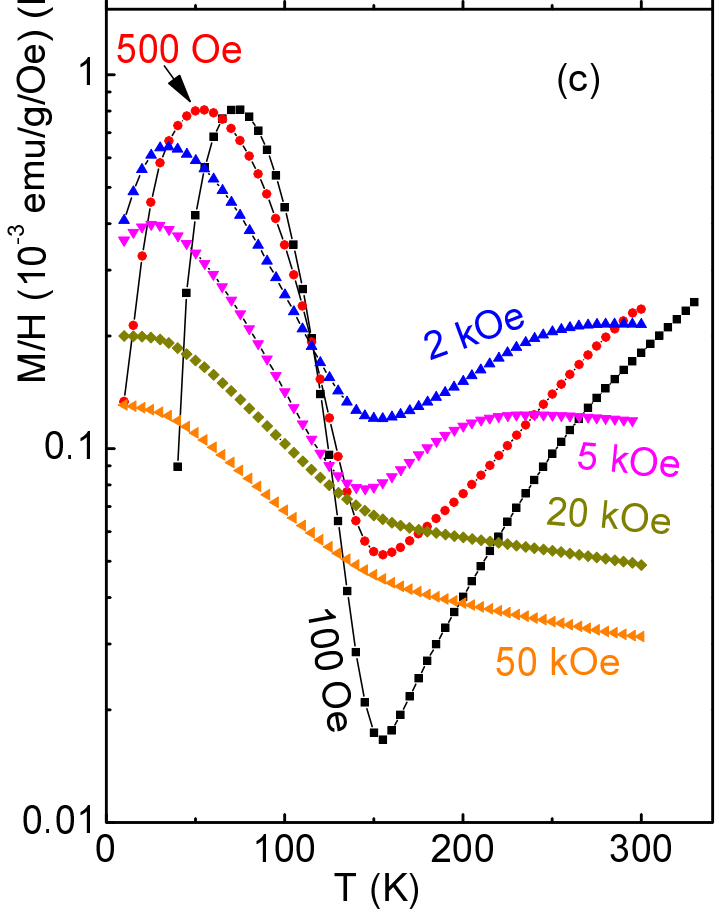
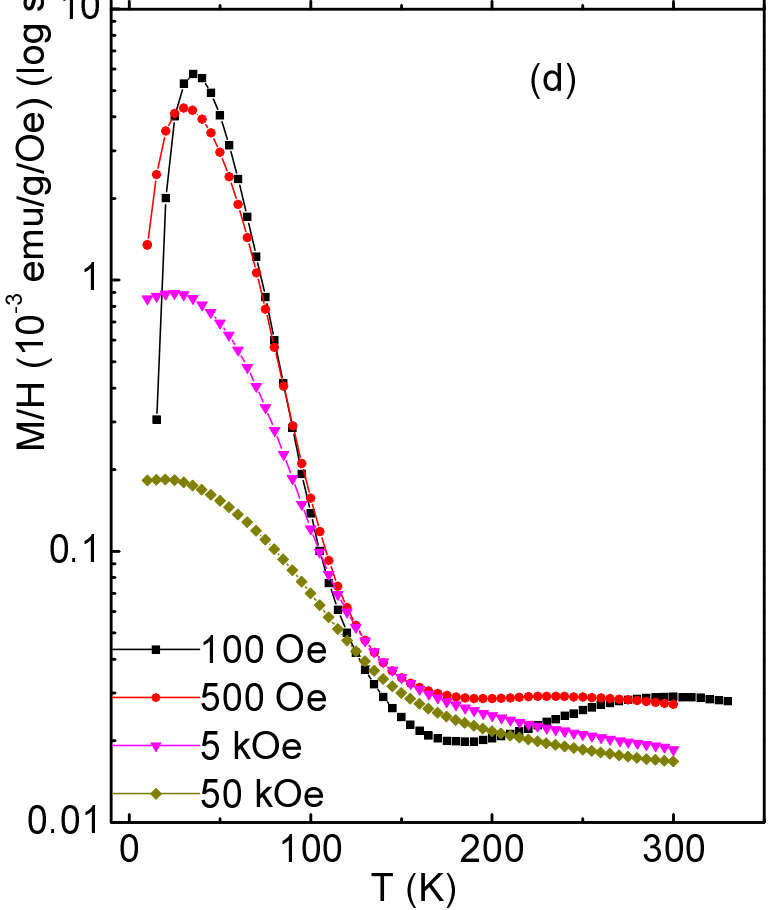

Figure 3. Temperature dependence of field normalized MZFC and MFC curves at 100 Oe (a-b) and MZFC curves at different fields (c-d) for the CF70 and CF90 samples. The inset of (b) shows a possible low temperature spin compensation effect at 100 Oe.

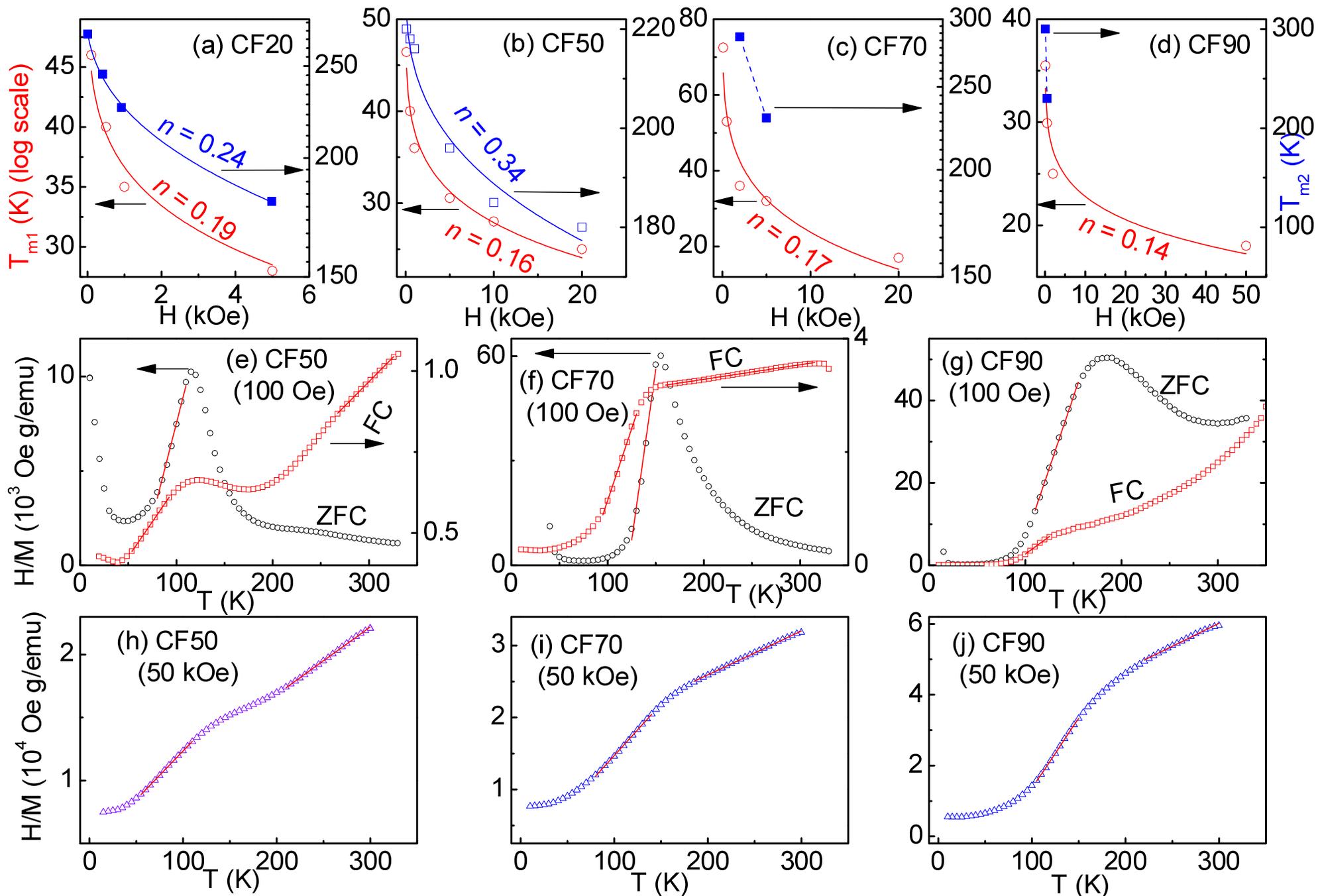

Figure 4. Field dependence of the peak temperatures ($T_{m1}$: left Y axis and $T_{m2}$: right Y-axis), along with power law fit and exponnet values (a-d), temperature variation of the inverse of dc susceptibility in ZFC and FC modes, along with Curie-Weiss law fit for lower field at 100 Oe(e-g) and for higher field at 50 kOe in ZFC mode (h-j) of the samples.

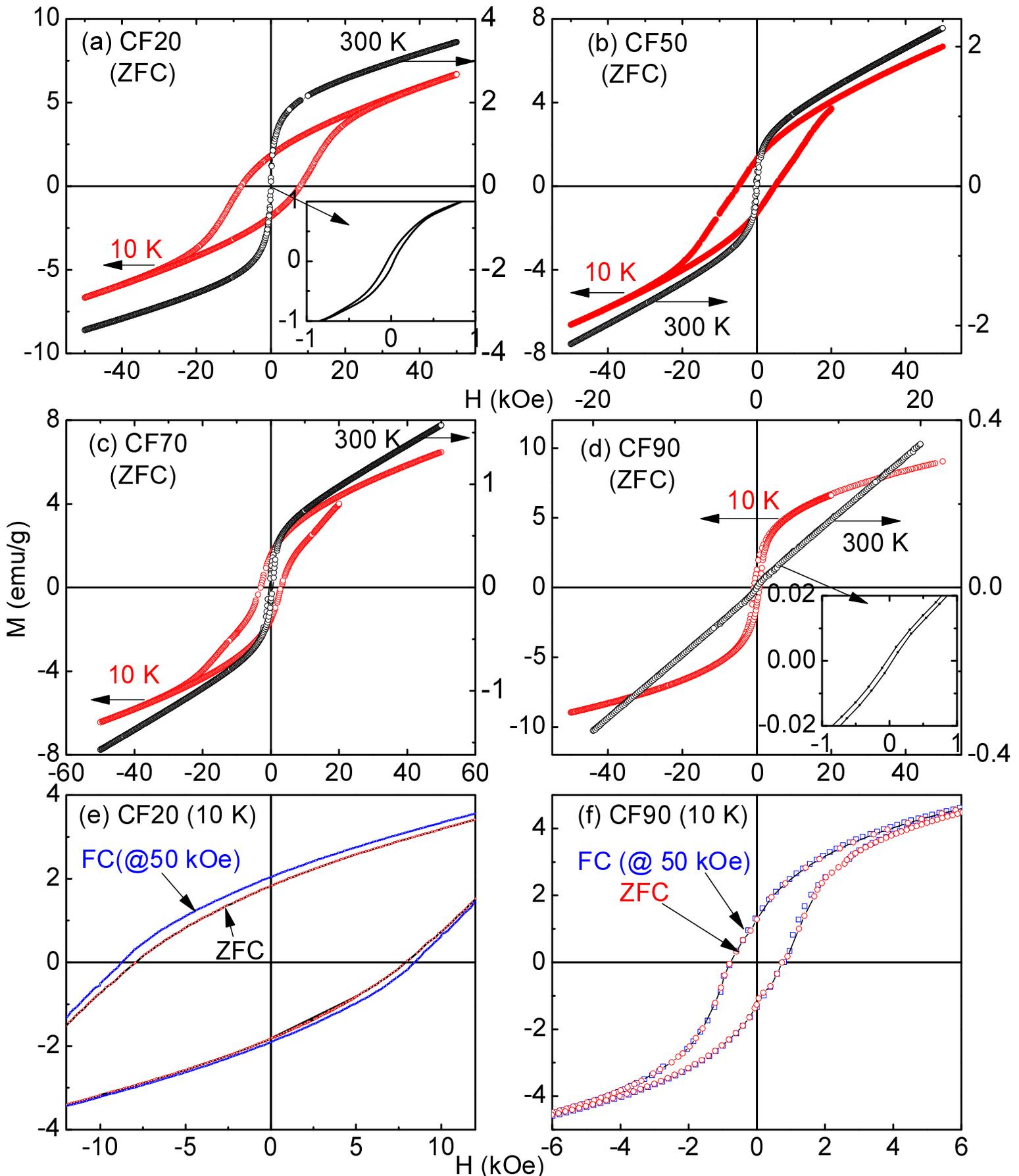

Figure 5 Field dependence of the magnetization curves at 10 K and 300 K for different samples (a-d) and the inset of (a, d) confirms the existence of loop at 300 K. The FC and ZFC loop at 10 K are compared in (e-f) for two samples.

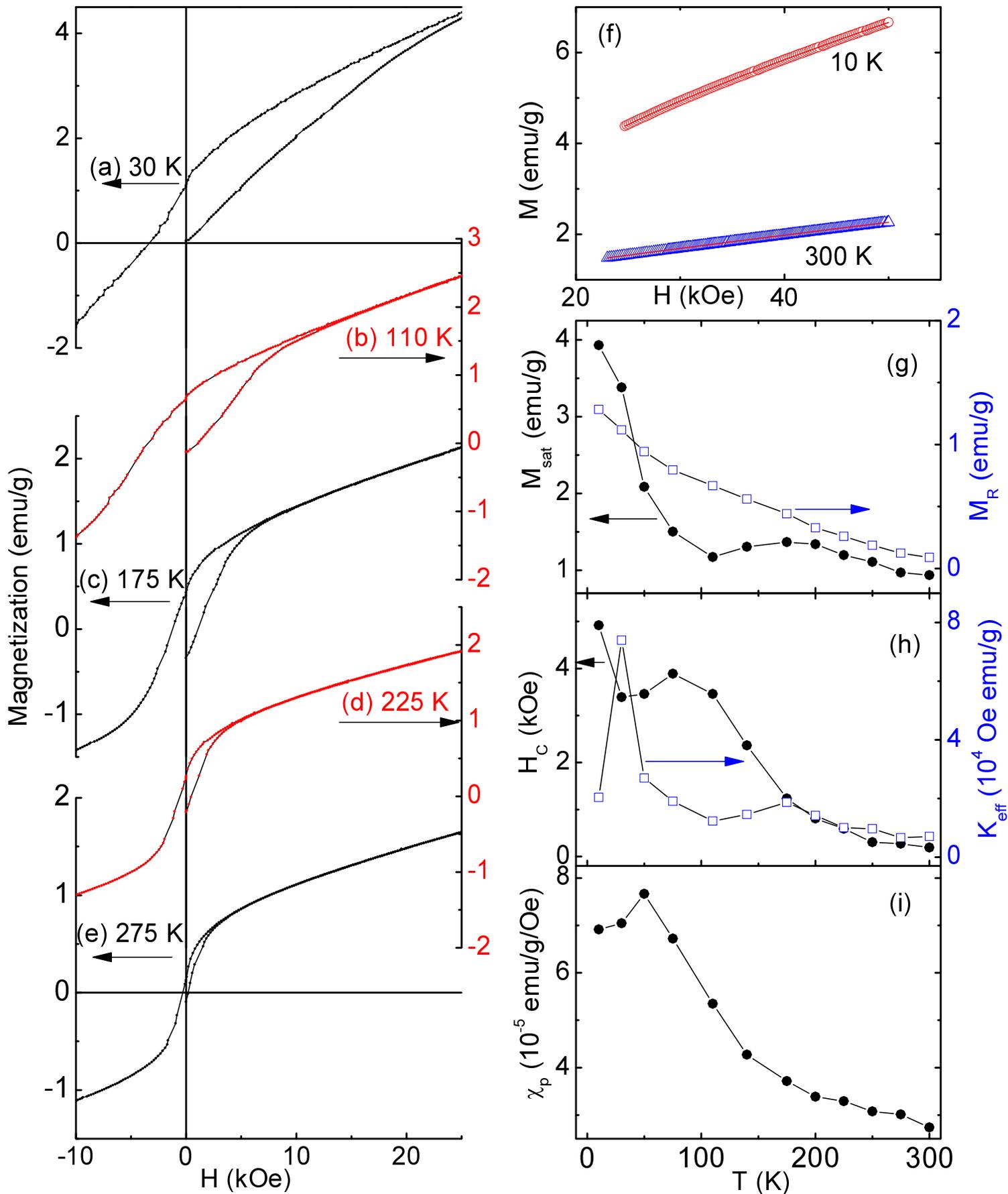

Figure 6. M(H) curves of the CF50 samples at selected temperatures (a-e). The fit of initiial M(H) curve at 10 K and 300 K are shown in (f). The temperature variation of the magnetic parameters from M(H) curves are shown in (g-i).

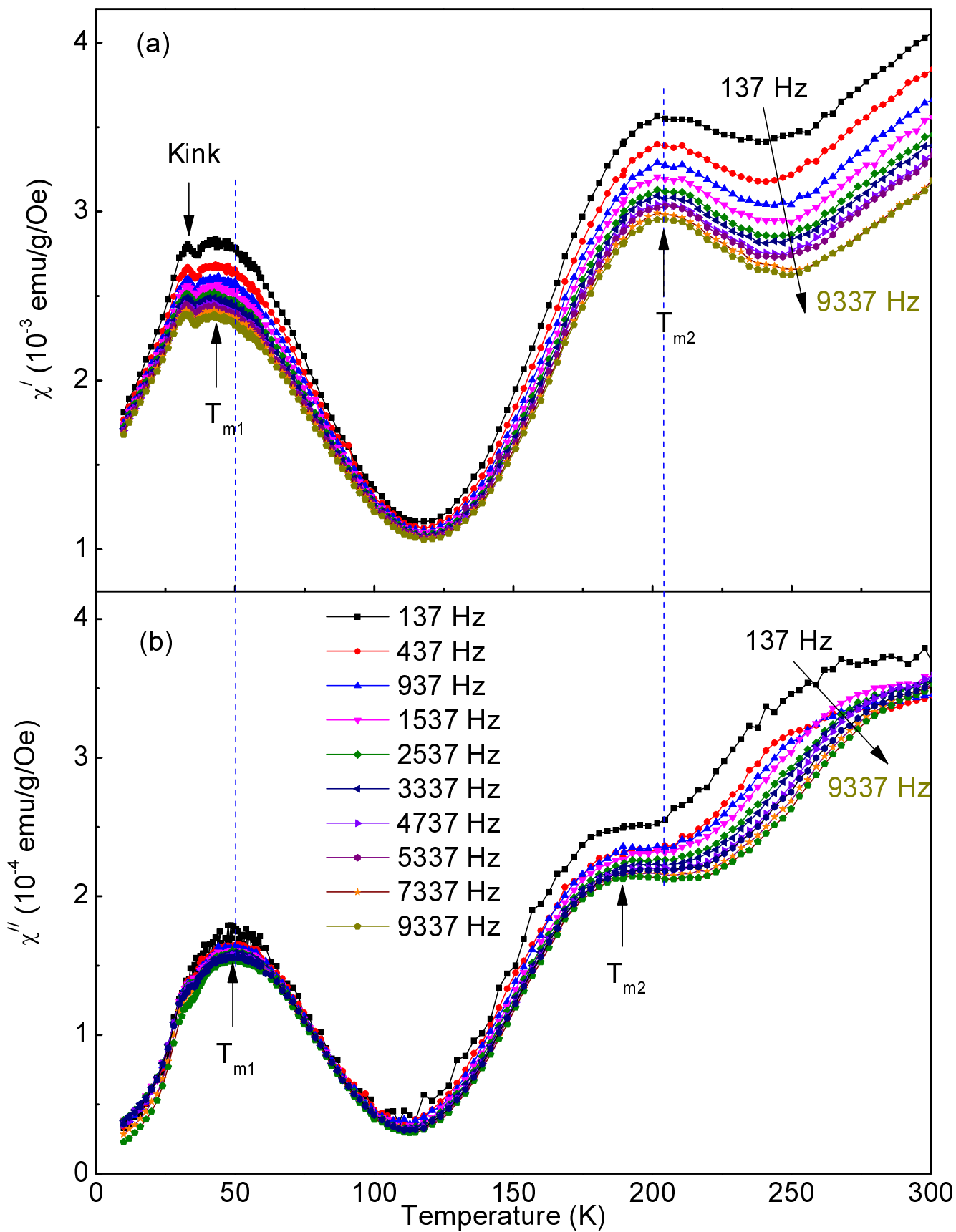

Figure 7 The temperature dependence of real (a) and imaginary (b) components of ac susceptibility of CF50 sample, measured at driven frequency in the range 137-9337 Hz. The difference in peak position of the real and imaginary components are shown by vertical dotted lines, where as the peak positions are shown by vertical arrows..

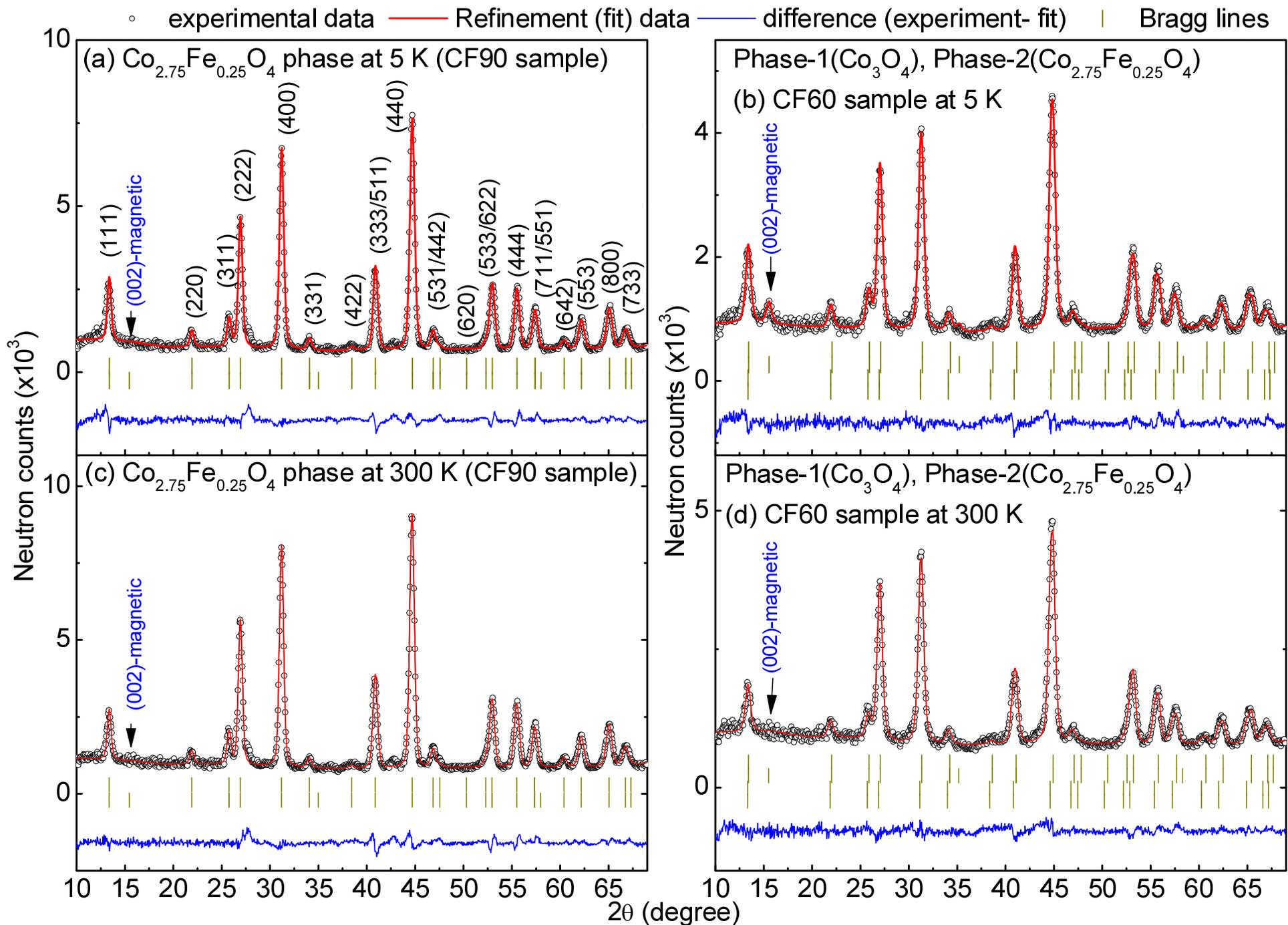

Fig. 8 Rietveld refinemnt of the ND patterns for CF60 and CF90 samples at 5 K and 300 K (a-d). The bragg peaks of the cubic spinel stucture are indexed for the ND pattern of CF90 sample at 5 K (a).

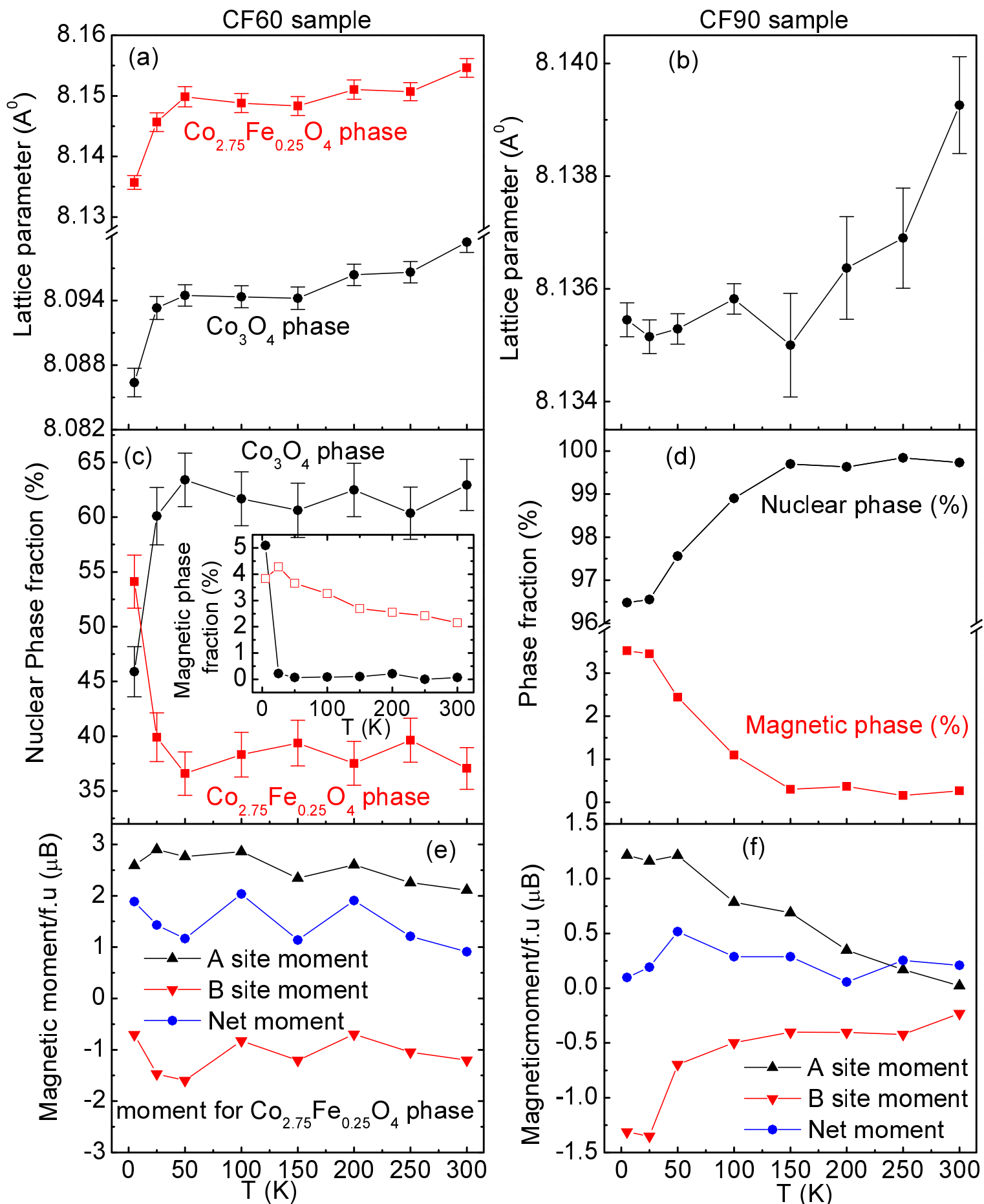

Fig.9 Temperature variation of the lattice parameter (a, b), phase fraction (c, d) and magnetic moment distribution per formula unit of cubic spinel structure (e, f).

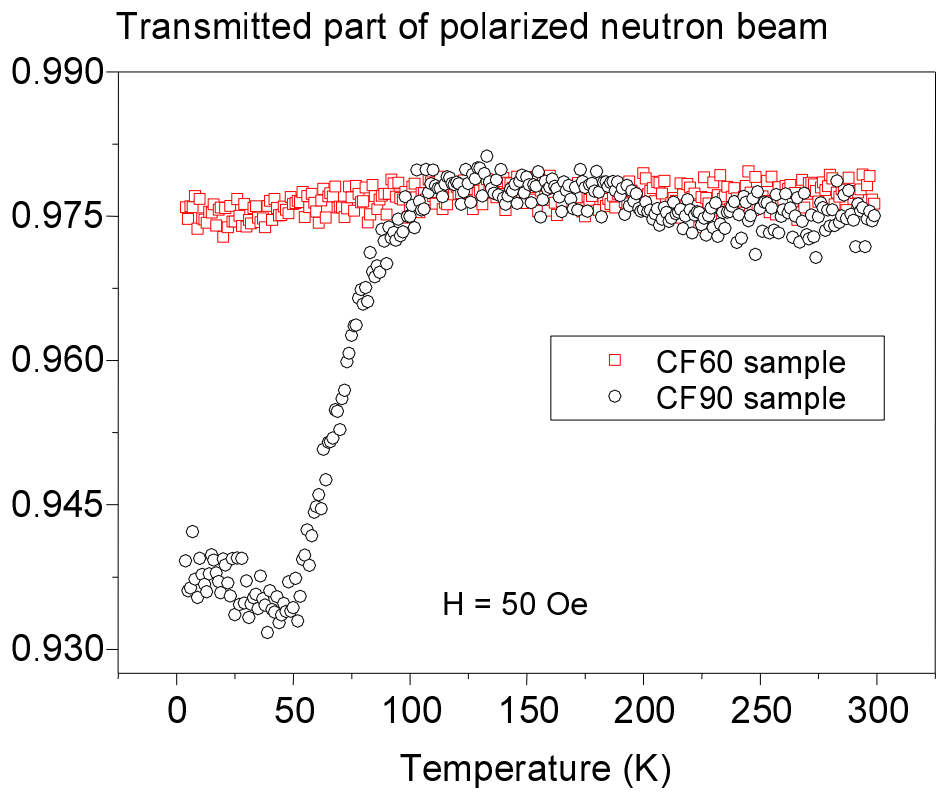

Figure 10. Neutron depolarization curves for CF60 and CF90 samples.